\documentclass[12pt,a4paper]{article}
 
\usepackage{epsfig}
\def\lb{\left(}
\def\rb{\right)}
\begin{document}
\setcounter{page}{0}
\thispagestyle{empty}

\large
\centerline{\bf Nanoscale Defect Formation on InP(111)}
\vspace*{-0.2cm}
\centerline{\bf  Surfaces after MeV Sb Implantation}
\normalsize
\vspace*{0.1cm}
\centerline {Dipak Paramanik$^{1}$, Asima Pradhan$^{2}$,and Shikha Varma$^{1,*}$}
\vspace*{0.1cm}
\small
\centerline{\it ${^1}$Institute of Physics, Bhubaneswar - 751005, India.}
\vspace*{-0.3cm}
\centerline{\it ${^2}$Department of Physics, IIT, Kanpur - 208016, India }
\normalsize
\vspace*{1.3cm}
\begin{center}
{\bf Abstract}
\vspace*{-.4cm}
\end{center}

We have studied the surface modifications as well as the surface roughness
of the InP(111) surfaces after 1.5 MeV Sb ion implantations. 
Scanning Probe Microscope (SPM) has been utilized to investigate the ion 
implanted InP(111) surfaces. We observe the formation of nanoscale defect
structures on the InP surface. The density, height and size of the 
nanostructures have been investigated here as a function of ion fluence. 
The rms surface roughness, of the
ion implanted InP surfaces, demonstrates two varied behaviors as a function of 
Sb ion fluence. Initially, the roughness increases with increasing 
fluence. However, after a critical fluence the roughness 
decreases with increasing fluence.  We have further applied the technique 
of Raman scattering to investigate the implantation induced 
modifications and disorder in InP. Raman Scattering 
results demonstrate that at the critical fluence, where the
decrease in surface roughness occurs, InP lattice becomes amorphous.

\vspace*{2.2cm}
\vspace*{0.2cm}
\noindent $^*$ Corresponding author: shikha@iopb.res.in; Tel:91-674-2301058; FAX:91-674-2300142
 
\normalsize
\newpage
\section{Introduction}
\noindent

InP is progressively being considered as a potential candidate 
for high speed electronic  devices and opto electronic applications 
\cite{str,hum,lam}. Sb is also considered an important dopant because
of its role in the development of field effect transistors and infrared 
detectors \cite{temp}. Due to its excellent physical properties like 
high thermal conductivity, high peak velocities for electrons and holes, 
InP is considered an important 
semiconductor material and it is being prominently utilized in the devices 
for high electron mobility transistors, high efficiency and high speed 
quantum well lasers, photo-detectors, photonic integrated circuits etc.
Although InP is being extensively used in semiconductor based 
device technology, it has been investigated much less compared 
to GaAs. Due to its low thermal stability, MeV ion 
implantation is a prominent way to introduce and dope the materials 
in InP. MeV implantation is also important for forming thick buried 
layers with modified properties as well as modification of vertically 
limited layers and quantum well structures.  
The increased density in VLSI circuits also makes 
the technological applications of the ion implantation, especially 
in MeV energy range, increasingly important. MeV implantation however 
can also produce severe modifications in the material depending on 
the nature and the energy of the impinging ion, and the implantation 
dose \cite{tam1}. 
Extensive usage of ion implantation in device fabrication and the 
continued miniaturization of device structures has brought the 
issue of surface modifications, via ion implantations, to the 
forefront. However, the factors responsible for such modifications 
and the surface morphology after ion implantation, have received 
little attention \cite{car}. Since roughness of surface
can crucially effect
the performance and reliability of devices \cite{dim}, it becomes 
necessary to characterize the surface roughness and understand the 
processes influencing it. The formation and the development of 
the surface structures, due to the ion implantation, are also gaining 
importance because of the realization that these structures can be 
utilized for controlled fabrication of semiconductors similar to 
self-organized growths. 

 Scanning Probe microscope (SPM) is a very effective tool for 
examining surface modifications and surface structures. 
There are very few studies in literature that have investigated 
the morphological changes of the ion implanted InP surfaces by SPM. 
Moreover, the surface studies on InP surfaces have been performed 
either after keV implantations \cite{yub} or after Swift Heavy Ion 
(SHI) irradiations at several hundreds of MeV \cite{kam,sin1}, and
there are no surface studies in literature where MeV implantations 
on InP have been investigated. In the present study we have made 
a detailed investigation on InP(111) surfaces after 1.5~MeV Sb 
implantation. 

During implantation, a projectile while moving forward produces
defects and loses energy due to scattering processes. The ion gets 
finally deposited at the range governed by its mass and implant energy 
\cite{chu}. At MeV energies, nuclear energy loss (S$_n$) processes are 
expected to be dominantly responsible for the material modifications. 
Defects and strains can get produced, via $S_n$, causing the
modifications in the surface and bulk properties \cite{gkm,sdey1}.  
Raman scattering intensity and peak shifts of the zone center 
phonon peak are very sensitive to these modifications.
Formation of defects can also lead to stress in the planes of 
the single crystal as well as changes in 
force constants. Corresponding shifts in the phonon frequencies 
are reflected in the Raman spectra. Our earlier study of MeV 
implantation in Si \cite{sdey2} had shown that the Raman scattering 
is a powerful technique for investigating 
and monitoring the implantation induced modifications.
It has also been shown to be a very sensitive probe for small 
volume defects created in Si
during ion implantation \cite{sdey2,hua,hua1}. Cusco et.al.  
have utilized Raman Spectroscopy to investigate the fluence dependant 
keV implantation of Si$^+$ in InP(100) \cite{cus}. Defect annealing 
after 2~MeV Yb$^+$ implantation, for a single fluence of 
1$\times 10^{13} cm^{-2}$ in InP(100) has also been investigated by 
Raman spectroscopy \cite{kat}. In the present study, in addition to SPM,
we have also utilized Raman scattering technique to investigate the fluence 
dependant modifications in InP after 1.5~MeV Sb implantation.
Moreover, the crystalline/amorphous phase transition of InP for high 
Sb fluences has also been observed here through Raman scattering. 

In this paper we have utilized the SPM technique to understand the 
modification in roughness and morphology of InP surfaces upon 
1.5~MeV Sb ion implantation. Formation of nanoscale defect structures 
on InP surface due to Sb implantation have been observed in 
SPM images. The height and size distributions of the nanoscale
defect structures have been presented here. We further 
observe that the MeV 
Sb ion implantation in InP leads to surface roughness that displays 
two different behaviours as a function of fluence. Initially, the 
surface roughness increases with increasing fluence. Beyond a critical 
fluence, however, the surface roughness decreases with increasing fluence.
We have utilized the Raman scattering results to understand
the modifications in surface roughness of InP after implantation. 
The Raman scattering results indicate that the critical fluence, where 
the surface roughness begins to decrease for increasing fluence, occurs 
at a stage when the InP lattice has become amorphous. The smoothening 
of the surface thus may be related to the amorphization in InP. 
Our results here are different than those observed after 100~MeV Au SHI 
(Swift Heavy ion) irradiation \cite{sin1} of InP(111) where, compared 
to our results, a 
lower rms surface roughness below 1$\times10^{13} ions/cm^2$ but a higher 
roughness for larger fluences was observed. 

Experimental procedures are discussed in section~2. Nanoscale defect 
formation on surfaces, rms surface roughness and Raman Scattering 
studies of InP(111) after implantation are presented in section~3. 
Conclusions are presented in section~4.

\section {Experimental}
\noindent     

A mirror polished (111)-oriented InP single crystal wafer was 
used in the present study. The samples were implanted at room temperature
with a scanned beam of 1.5 MeV Sb$^{2+}$ ions at various fluences
ranging from  1$\times 10^{11}$ to 5$\times10^{15} ions/cm^2$.
The average Sb flux was 0.02 $\mu$A/cm$^2$. This current was
measured directly on the target after suppressing the secondary
electrons by applying a negative bias of 200V to a suppressor
assembly around the target. The implantations were performed with
the samples oriented 7$^o$ off-normal to the incident beam to
avoid channeling effects. Monte Carlo simulations were performed 
for 1.5~MeV Sb implantation in InP using the SRIM'03 code and
the mean projected range of Sb-ion distribution
was found to be 400~nm \cite{srim}. 

Scanning Probe Microscope (SPM) Nanoscope IIIA from Veeco was 
used to image the implanted InP (111) sample surfaces with a 
silicon nitride cantilever operated in tapping mode. Images ranging 
from 0.2 to 10 $\mu{m}$ square were obtained. 
The root mean square (rms) surface roughness
was calculated by the SPM software.
 
Raman scattering measurements were performed using a SPEX 1877E
Triplemate Spectrometer with a liquid nitrogen cooled, charged coupled
device array. The laser power was controlled to avoid laser
annealing effect on the sample. Raman experiments were carried out
at room temperature using the 514~nm line of an Argon ion laser in
the backscattering geometry. At this wavelength the penetration depth
of the light is estimated to be about 100~nm. As the penetration depth 
of the light is much smaller than the projected range of the implanted 
Sb ions, Raman scattering results are primarily from the surface region.

\section {Results and Discussion}

Fig.1 shows the 10$\times 10\mu{m^2}$ 2D SPM images from the InP surfaces. 
The image from a virgin (un-implanted ) InP(111) sample is shown and
it is observed that this surface is smooth. Other images 
of Fig. 1 show the evolution of the surface morphology on InP surfaces 
after 1.5 MeV Sb implantation at fluences ranging from 
1$\times10^{11} ions/cm^2$ to 1$\times10^{15} ions/cm^2$. Comparison
of the surface morphologies of InP surfaces of Fig.~1, after implantation, 
show the formation of nanoscale sized defects with varying size, height 
and density depending on the 
fluence.  Fig.1a shows the InP surface after an Sb implantation with 
1$\times10^{11} ions/cm^2$. Several nano sized defects can be observed 
on the surface. The structures have developed due to the damage created 
at the surface. Some surface structures have been earlier reported 
after  keV Ar ion irradiation of InP surfaces \cite{yub} and 
4.5 MeV Au implantation on HOPG surfaces \cite{wan}. Nanoscale defects on
InP surfaces have also been observed after 100~MeV Au SHI irradiations
\cite{sin2}. However, the present study is the first study of modifications
of InP surfaces after MeV implantation. We have investigated 
the height and the size distribution of the nanoscale sized defects 
(seen in Fig.~1) on the InP surfaces after various Sb fluences. The 
size and the height
distributions are shown in Fig.2 and 3 respectively. After a fluence of 
1$\times10^{11} ions/cm^2$, most of the nanostructures have a diameter 
smaller than 450~nm and a height smaller than 10~nm. 
The density of the nanostructures has been calculated 
to be about 2.5$\times 10^{8}cm^{-2}$.  Fig.1b shows the InP surface 
image after an Sb fluence of 1$\times10^{12} ions/cm^2$. We observe that
the nanostructures have become bigger in size. As seen in the size distribution
of Fig.~2b, some structures have diameter as large as 1200nm. However, a 
large number of nanostructures have diameter smaller than 200nm. Although 
some nanostructures are as high as 18~nm, most of the nanostructures are 
lower than 12~nm (Fig.~3b). Furthermore, a large number of nanostructure have 
a height lower than 4~nm. The total density of the nanostructures is found to 
be similar to that observed at 1$\times10^{11} ions/cm^2$. At
1$\times10^{13} ions/cm^2$, in Fig.1c, we notice a slight increase in the 
density of nanostructures to 3.6$\times10^{8}cm^{-2}$. A few nanostructures 
have diameters as large as 950~nm. A large number of nanostructures, greater
than at 1$\times10^{12} ions/cm^2$, have diameters smaller than 200~nm 
(Fig.~2c). Although some are 20~nm high, a large number have a height lower 
than 12~nm (Fig.~3c). Again, a large number of nanostructures have a height 
lower than 4~nm. The increase in density at this stage can also be noticed 
by a changed (y) scale for both the distributions. Fig.~1d shows the 
image acquired 
after the fluence of 1$\times10^{14} ions/cm^2$. The density of the 
nanostructures is about 5.0$\times10^{8}cm^{-2}$. The size 
and the height distribution is very similar to that observed at 
1$\times10^{13} ions/cm^2$. However, the diameter of the largest 
nanostructures observed is smaller (700~nm) and the number of small 
(diameter less than 100~nm) nanostructures has increased (Fig.~2d). Also, 
larger number have a height lower than ~4nm (Fig.~3d). After a fluence of 
5$\times10^{14} ions/cm^2$, a drastic increase in density of the 
nanostructures is observed in Fig.~1e. We also observe a larger number 
of nanostructure with small size. The density of nanostructures at this 
stage is about 8.0$\times10^{8}cm^{-2}$. Although the size distribution is 
similar to that observed at 1$\times10^{14} ions/cm^2$, there are many 
more nano-structures with small 0-100~nm diameter (Fig.~2e). Similarly the 
nanostructures having height smaller than 4~nm has increased (Fig.~3e). 
The SPM image after a fluence of 1$\times10^{15} ions/cm^2$ is shown in
Fig.~1f. The density of the nanostructures, 8.0$\times10^{8}cm^{-2}$, 
as well as the size and the height distributions are very similar to 
those observed after 5$\times10^{14} ions/cm^2$. However, some 
structures of large 1000~nm diameter (Fig.~2f) are also seen.
Similar size and height distributions were also observed for 
5$\times10^{15} ions/cm^2$. Here we notice that for all fluences, the
defect density is far lower than the ion beam fluence. Possible reasons 
for this will be discussed below.

Figure~4 shows the high resolution $1 \times 1\mu{m^2}$, 
$0.5 \times 0.5\mu{m^2}$ and $0.2 \times 0.2\mu{m^2}$ images 
of the InP(111) surfaces after the fluence of 
$1\times10^{13} ions/cm^{2}$ and $5\times10^{14} ions/cm^{2}$. 
The figures show that the InP(111) surfaces are very different, 
at each scale, for these two fluences. Some SPM -sections with 
images are shown in Fig.~5 to display some characteristic features 
of the distribution of the defects on InP surfaces. Fig.~5a shows 
a $1.0 \times 1.0\mu{m^2}$ SPM image of InP surface after an ion 
fluence of $1\times10^{11} ions/cm^{2}$ along with the section analysis 
demonstrating a defect of 78.1~nm lateral and 1.8~nm vertical dimensions.
Some defects of smaller size and height are also visible. Fig~5b
shows $1.0 \times 1.0\mu{m^2}$ SPM image of InP surface after an ion 
fluence of $1\times10^{12} ions/cm^{2}$. Some small and big sized 
defects are visible. Section analysis of a defect with dimensions 534~nm
in lateral and 2.5~nm in vertical direction is also shown. We interestingly
notice that this defect is actually composed of several smaller defects.
These features can be clearly seen in Fig.~5c where a high resolution   
$0.4 \times 0.4\mu{m^2}$ SPM image of this defect (from Fig.5b) is shown.
To emphasize, image in Fig.~5c shows the internal structure of the big 
defect analyzed in fig.~5b. As seen in Fig.~5c,
the smaller defects embedded in the big defect are of several sizes and 
heights. Section-analysis of a typical small defect is shown in Fig.5c
with dimensions of 34.0~nm in lateral and 1.0~nm in the vertical direction.
The image also shows that several defects are overlapping other defects.
To our knowledge, these kind of high resolution SPM images of the ion beam 
induced defects have never been reported in the literature.  
Fig.~5d shows a $1.0 \times 1.0\mu{m^2}$ SPM image for a fluence of 
$1\times10^{13} ions/cm^{2}$. The section analysis shows a defect
of 62.5~nm lateral and 4.9~nm vertical dimensions. Again, the big
defects clearly appear to be composed of several smaller defects.
Several small and lower defects can also be seen spread over 
the surface. Similar behaviour is also noticed in Fig.5e where a   
$1.0 \times 1.0\mu{m^2}$ SPM image is shown for a fluence of 
$5\times10^{14} ions/cm^{2}$. The bigger defects are fully embedded 
with several smaller defects of various heights. The section analysis 
shows a defect with 46.9~nm lateral and 1.7~nm vertical dimensions. A 
big defect is shown in a $0.5 \times 0.5\mu{m^2}$ image of Fig.5f for 
a fluence of $1\times10^{15} ions/cm^{2}$. Here also the big defect is 
embedded with several smaller defects. The section analysis shows a defect with
27.0~nm lateral and 1.0~nm vertical dimensions. Several smaller defects 
can also be seen around the big defect. These images show that the bigger
defects at all fluences are embedded with several of nanosized defects. 

During the investigation of  $10 \times 10\mu{m^2}$ images of Fig.1,
we had noticed that the density of defects for all fluences varies
between 2.5-8.0$\times10^{8}cm^{-2}$, which is much lower compared 
to the ion fluences. From fig.~5, we notice that 
the bigger defect structures are composed of smaller nanosized-defects 
of sizes $\sim$ 30~nm. Taking this fact into account we have recalculated
the density of defects and find it to be 5.0$\times10^{10}cm^{-2}$  
at 1$\times10^{11} ions/cm^2$ and 1 - 1.5$\times10^{11}cm^{-2}$  for 
higher fluences. We further
notice, in Figs.~3 and 5, that the height of defects also vary as a function
of fluence. Although, most of the defects are about 4~nm high, 
higher defects are also increasingly seen at larger fluences (see Fig.~3).
In the framework of model introduced by Gibbons\cite{gib}, the amorphous
material is produced either directly by a single incoming ion or 
by multiple overlaps.  According to this model, the ratio between the 
total surface area A$_A$ covered by damages and the total area A$_0$ 
being implanted is given by

\vspace*{6mm}

$$ {A_A \over A_0} = {1- e^{-A_1 \phi}}  \Sigma_{k=0}^m  
{{\lb \displaystyle {A_1 \phi}\rb} ^k \over\displaystyle k!}$$

\vspace*{6mm}

where A$_1$ = $\pi {r_m}^2$ is the surface area damaged by a single
ion impact, $\phi$ is the fluence and {\it m} is the overlap number. 
For $1\times10^{11} ions/cm^{2}$ with $r_m$ of 30~nm
we find that {\it m} is 2, i.e. about two ions must impinge on 
the same area to produce the defect. For 250~MeV Xe irradiation of InP, 
Herre et al. find that the values of {\it m} is between 2 and 3 \cite{her}.
Higher heights of defects for larger fluences, as observed in Fig.~3, 
may denote larger {\it m}.
In addition, more than one defect may be getting formed at one place. 
Overlapping defects as well as defects smaller than 30~nm
have also been seen in Fig.~5. All these factors 
together can be responsible for the observation of the lower defect 
density than the ion fluences.

We have also studied the rms surface roughness of the InP surfaces
after MeV ion implantation. In Fig.~6 we have plotted the rms surface 
roughness($\sigma$) of the InP surfaces as a function of ion fluence.
For a virgin InP(111) surface, $\sigma$ was measured to be 0.47~nm 
and is also marked in Fig.~6. We observe that the rms surface 
roughness exhibits two distinct behaviors as a function of fluence. 
Initially up to $1\times10^{14} ions/cm^{2}$, $\sigma$ increases
with the increasing fluence. However for higher fluences 
$\sigma$ decreases for increasing fluences. Our results show
that there is a critical fluence of $1\times10^{14} ions/cm^{2}$,
below which the rms roughness of the InP surfaces increases with ion
fluence whereas for higher fluences the surface roughness decreases 
with increasing fluences.  A similar decrease 
in surface roughness with increasing fluence, beyond a critical 
fluence, has been observed for MeV Sb implantation in Si(100)
\cite {sdey5} and for keV implantations of P and As 
in amorphous films \cite{edr}. Comparing results of
100~MeV Au SHI irradiation on InP (S$_n$ = 378~eV/nm, 
S$_e$ = 15~keV/nm) studies with our results (S$_n$ = 2~keV/nm, 
S$_e$ = 1~keV/nm) we expect a higher surface rms roughness
at all fluences here as the  S$_n$ is higher \cite{gkm}. 
Although for fluences upto $1\times10^{13} ions/cm^{2}$ this is
seen, for $1\times10^{14} ions/cm^{2}$ we observe a lower roughness
in our case. This is an unexpected result and suggests that at 
higher fluences factors other than S$_n$ are also playing role.
At high fluences, density of electronic excitations increase, covalent
bonds in the lattice weaken or get broken. As a result the lattice 
softens. This softening of the bonds and amorphization of the InP
lattice has been shown by our Raman scattering results discussed below.
The SHI studies \cite{sin1} did not investigate the fluences higher than
$1\times10^{14} ions/cm^{2}$ and also did not observe any decrease in the
roughness. Our Raman scattering results, presented next, indicate the 
occurrence of amorphization in InP at this fluence.

Fig. 7 shows the as-implanted Raman spectra from the InP samples 
implanted with various Sb doses. All these spectra were acquired 
in the backscattering geometry. The spectrum from a virgin 
(unimplanted) InP is also shown for comparison. The spectra have 
been shifted vertically for clarity, but the intensity scale is the 
same for all the spectra. The spectrum of the virgin InP (Fig.7) shows 
the characteristic longitudinal optical (LO) and transverse optical 
(TO) Raman peaks of crystalline InP(111) 
\cite{pin}. The features at 305~cm$^{-1}$ and at 347~cm$^{-1}$ are 
assigned to the TO and the 
LO phonon modes, respectively. The sequence of spectra gradually
evolve, with increasing fluence, from the characteristic crystalline
InP(111)  spectrum  to the amorphous like spectrum  of Fig.~7f.
The spectrum for the 1$\times10^{11} ions/cm^2$ sample (Fig.~7a)
exhibits some changes when compared to virgin InP. We observe that
in addition to the shifts of both LO and TO features towards the
lower wave numbers, TO feature also exhibits an asymmetric broadening
towards the lower wave numbers. All these changes reflect
the modifications in the InP due to the defects created during
implantations. After a fluence 1$\times10^{12} ions/cm^2$  (Fig.~7b) 
we observe a decrease in the intensity of the TO mode. In addition,
broadening as well as the shifts towards lower wave numbers
are observed, for both LO and TO modes. 
Spatial Correlation model related to {\it q}-vector relaxation induced 
damage shows \cite{tio} that when disorder is introduced into the crystal 
lattice by implantation, the correlation function of the phonon-vibrational 
modes becomes finite due to the induced defects and consequently the 
momentum {\it {q=0}}  selection rule is relaxed. Consequently, the phonon 
modes shift qualitatively to lower frequencies and broaden 
asymmetrically as the ion fluence is increased \cite{ric}.  
Thus the shifts to lower frequencies as well as the asymmetrical 
broadening of the features, observed in Fig.~7, are due to the residual 
defects created via implantation.
Accordingly, these two features are also referred to as DALO and
DATO respectively for disorder activated (LO) and (TO) modes. 
The shifts, of the LO and TO modes, towards the lower wave numbers 
also indicate the development of the tensile strain in the lattice.
Our results are in contrast to the studies of Si$^+$ implantation
in InP at 150 keV \cite{cus} where no noticeable changes  compared 
to the virgin InP were seen upto the fluence of 1$\times10^{12} ions/cm^2$
and the first signatures of disorder were observed after the fluence of 
5$\times10^{12} ions/cm^2$. For 2~MeV Se implantation in InP, however,
some damage after 1$\times10^{12} ions/cm^2$ has been reported using
channeling experiments\cite{wes}.

In Fig.7c after a fluence of 1$\times10^{13} ions/cm^2$, a further 
decrease in the TO mode intensity as well as increased 
broadening and shifts of LO, TO modes towards smaller wave 
numbers are observed. 
After a fluence of 1$\times10^{14} ions/cm^2$, the Raman spectrum 
(Fig. 7d) exhibits no distinct features corresponding to LO or the TO 
modes indicating that at this stage the lattice has been amorphised. The
DATO and DALO structures have become completely merged into a broad 
band containing the whole density of states of the optical modes.
This spectrum resembles that of amorphous InP \cite{wih}. Hence, we notice 
that InP has become amorphised at 1$\times10^{14} ions/cm^2$ and 
further increase of fluences does not  produce changes in the
LO or TO modes (Fig.~7e, 7f). Since the penetration depth of Ar$^+$ 
laser is 100~nm in InP, the Raman results here are primarily from 
surface region. The fluence, 1$\times10^{14} ions/cm^2$, where InP 
becomes amorphised is surprisingly similar to that observed 
at keV energies \cite{cus,ken} or even at SHI energies \cite{her}.   
The decrease in rms surface roughness, $\sigma$, (in Fig.~6) can 
thus be related to the amorphization of the InP at this fluence.
The amorphization can lead to relaxations \cite{vol,sri} and 
smoothening of the surface via decreased strains \cite{sdey2}.

   There are some experimental observations on the 
nucleation and growth of defects formed by ion implantation 
in crystalline InP at keV \cite{yub} and at SHI \cite{sin1,sin2} 
energies. However, such studies are not present at MeV scales. 
The evolution of the surface morphology during ion 
bombardment will be governed by a balance between the roughening 
and the smoothening processes. The random arrival of the ions on 
the surface constitutes the stochastic surface roughening.
Surface diffusion, viscous flow and surface sputtering etc. 
contribute towards the smoothening of the surface \cite{ekl}. 
The mechanism for the formation of surface damage is also postulated
as a result of cascade collisions due to nuclear energy loss.
S$_n$ has been considered to be mainly responsible for the
surface modifications of InP after 2~MeV Se implantation \cite{wes} 
and high energy ($\sim$100 MeV) Au irradiation \cite{sin1}. 
In the present study also S$_n$ seems to be the dominating factor 
in the creation of the nanostructure after Sb implantation.
The 4.5~MeV Au implantation in HOPG \cite{wan} results in
protrusions and features similar to the nanostructures seen here 
on InP surfaces. The mechanism for the formation of the surface 
features, on HOPG, was also S$_n$ dominated. 
The nano-sized structures observed here, after MeV implantation, 
are smaller in size compared to structures seen after 2~keV Ar 
irradiation on InP surfaces where size ranged between 30-60~nm. 
Moreover, the height of the structures was over 100~nm after keV 
irradiation whereas it is always lower than 18~nm, and mostly
around 4~nm, in the present case. Differential sputtering of
InP surfaces leading to In-rich zones was suggested to be a possibility
for the nucleation of surface structures after keV irradiation \cite{yub}.
Similar scenario may be taking place at MeV energies also. Thus,
S$_n$ related processes, differential sputtering of a component,
and the presence of tensile stress as observed in Raman spectra by 
softening of LO, TO modes, may be all together responsible for 
creating the nano-sized defects observed here after MeV Sb implantations.

\section{Summary and conclusions}
\noindent

In the present study, the modifications in the surface morphology of 
InP(111) have been examined after 1.5~MeV Sb implantation. The InP 
surfaces display nano-sized defect structures. The height and size 
distributions of the nanostructures have been studied here. 
For fluences of 1$\times10^{12}$ - 1$\times10^{15}ions/cm^2$,
several nanostructures of sizes smaller than 100~nm and lower than 
4~nm have been observed. Larger and bigger defects are observed to 
be embedded with smaller nano-sized defects in the SPM images.
The surface roughness initially increases upto 
the Sb fluence of 1$\times10^{14}ions/cm^2$. For higher fluences a 
decrease in surface roughness is observed. Raman Scattering results 
indicate that InP becomes amorphous at this stage. The decrease in 
surface roughness is related to the smoothening of surface due to 
amorphization. Nuclear energy loss processes and the presence
of tensile stress, as shown by the softening of LO,TO modes by Raman
scattering, may be together responsible for the formation of
the nanoscale defect structures on the InP surfaces.  

\section{Acknowledgments}
\noindent

This work is partly supported by ONR grant no. N00014-97-1-0991.
We would like the thanks N.C. Mishra for useful discussions.

\newpage
\noindent{\bf \Large {Figures}}

\vskip 0.3 in
\noindent Fig. 1: $10\times 10\mu{m^2}$ SPM images of InP 
surfaces for the virgin sample as well as after implantation 
with 1.5 MeV Sb ions at a fluence of (a) $1\times 10^{11} ions/cm^2$, 
(b) $1\times 10^{12} ions/cm^2$, (c) $1\times 10^{13} ions/cm^2$, 
(d) $1\times 10^{14} ions/cm^2$, (e) $5\times 10^{14} ions/cm^2$.
and  (f) $1\times 10^{15} ions/cm^2$ 

\vskip 0.3 in
\noindent Fig. 2: Size distributions of the surface structures 
after 1.5 MeV Sb implantation with fluences of 
(a) $1\times 10^{11} ions/cm^2$, 
(b) $1\times 10^{12} ions/cm^2$, (c) $1\times 10^{13} ions/cm^2$, 
(d) $1\times 10^{14} ions/cm^2$, (e) $5\times 10^{14} ions/cm^2$.
and  (f) $1\times 10^{15} ions/cm^2$

\vskip 0.3 in
\noindent Fig. 3: Height distributions of the surface structures 
after 1.5 MeV Sb implantation with fluences of 
(a) $1\times 10^{11} ions/cm^2$, 
(b) $1\times 10^{12} ions/cm^2$, (c) $1\times 10^{13} ions/cm^2$, 
(d) $1\times 10^{14} ions/cm^2$, (e) $5\times 10^{14} ions/cm^2$.
and  (f) $1\times 10^{15} ions/cm^2$

\vskip 0.3 in
\noindent Fig. 4: InP surface SPM images (a) $1\times 1\mu{m^2}$, 
(b)$0.5\times 0.5\mu{m^2}$ and (c) $0.2\times 0.2\mu{m^2}$ after
implantation at fluence of $1\times 10^{13} ions/cm^2$. Images 
(d) $1\times 1\mu{m^2}$, (e)$0.5\times 0.5\mu{m^2}$ and 
(f) $0.2\times 0.2\mu{m^2}$ are after implantation at fluence 
of $5\times 10^{14} ions/cm^2$.

\vskip 0.3 in
\noindent Fig. 5: InP surface SPM images and SPM-section
analysis of (a) $1\times 1\mu{m^2}$ image for $1\times 10^{11} ions/cm^2$
(b) $1.0\times 1.0\mu{m^2}$ image for $1\times 10^{12} ions/cm^2$
(c) $0.4\times 0.4\mu{m^2}$ image for $1\times 10^{12} ions/cm^2$
(d) $1\times 1\mu{m^2}$ image for $1\times 10^{13} ions/cm^2$
(e) $1\times 1\mu{m^2}$ image for $5\times 10^{14} ions/cm^2$
(f) $0.5\times 0.5\mu{m^2}$ image for $1\times 10^{15} ions/cm^2$.
(L is the lateral dimension and H is the height of the nanostructure 
labelled with arrows)

\vskip 0.3 in
\noindent Fig. 6: The rms surface roughness ($\sigma$) of the Sb 
implanted InP(111) surfaces, measured using SPM, is plotted as a 
function of Sb ion fluence.  Data for the virgin sample is also shown. 

\vskip 0.3 in 
\noindent Fig. 7: : Raman spectra are shown for virgin InP(111)
as well as after 1.5 MeV Sb implantation of InP with various fluences
of (a) $1\times 10^{11}$, (b) $1\times 10^{12}$, (c) $1\times 10^{13}$ 
(d) $1\times 10^{14}$, (e) $5\times 10^{14}$ and 
(f) $1\times 10^{15} ions/cm^2$. 
The curves are vertically displaced for clarity.

\end{document}